\documentclass[12pt]{article}
\usepackage{amssymb}
\usepackage{graphicx}
\usepackage{amsmath,amscd}
\def\D{\mathcal{D}}
\def\L{{\cal L}}
\def\Tr{{\rm Tr}}

\def\1{\amssymb{1}}
\def\E{\mathcal{E}}

\newcommand {\Wb}{\bar{W}}
\newcommand {\parb}{\bar{\partial}}
\newcommand {\Nf}{N_{\rm F}} 
\newcommand {\Nc}{N_{\rm C}} 
\makeatletter
\@addtoreset{equation}{section}

\makeatother

\makeatletter
\newcommand{\thetablename}{Table}
\def\fnum@table{\thetablename\ \thetable}
\makeatother
\begin{document}

\thispagestyle{empty}
\begin{flushright}

{\tt hep-th/0605064} \\
May, 2006 \\
\end{flushright}
\vspace{3mm}

%%%%%%%%%%%%%%%%%%%%%%%%%%%%%%%%%%%%%%%
\begin{center}
{\huge Non-integrability \\of Self-dual Yang-Mills-Higgs System}
\end{center}

%%%%%%%%%%%%%%%%%%%%%%%%%%%%%%%%%%%%%%%%%

\begin{center}
\lineskip .45em
\vskip1.5cm
{\large Takeo Inami$^1$\footnote{E-mail: inami@phys.chuo-u.ac.jp}, 
Shie Minakami$^1$\footnote{E-mail: shiye@phys.chuo-u.ac.jp} and
Muneto Nitta$^2$\footnote{E-mail: nitta@phys-h.keio.ac.jp}}
\vskip 1.5em
{\large\itshape $^1$Department of Physics, Chuo University\\
\itshape Kasuga, Bunkyo-ku, Tokyo, 112-8551, Japan.

\itshape $^2$ Department of Physics, Keio University\\
\itshape Hiyoshi, Yokohama, %Kanagawa 
223-8521, Japan. 

}  \vskip 4.5em
\end{center}

\begin{abstract}
We examine integrability of 
self-dual Yang-Mills system in the Higgs phase, 
with taking simpler cases of vortices and domain walls. 
We show that the vortex equations and the domain-wall equations 
do not have Painlev\'e property. 
This fact suggests that these equations are not integrable. 

\end{abstract}

%%%%%%%%%%%%%%%%%%%%%%%%%%%%%%%%%%%%%%%%%%%%%%%%%%%%%%

%%%%%%%%%%%%%%%%%%%%%%%%   1   %%%%%%%%%%%%%%%%%%%%%%%%%%
\newpage

\section{Introduction}

Field theories in low dimensions as well as gauge theories (in Higgs phase in some cases) in $d=3+1$ and other dimensions allow various kinds of solitons and soliton-like objects \cite{'tHooft:1974qc}--\cite{ANO}, 
see \cite{MantonSutcliffe}. 
Study of such objects, 
particularly those of BPS class \cite{Bogomolny:1975de}, 
has become a subject of growing interest in view of applications 
to particle physics and mathematical physics. 
One interesting recent development is the relationship of such objects of different dimensionality in gauge theories. 
This view was extended to a few other cases in 
susy gauge theories. 
Another development is the discovery of 1/4 BPS composites 
of such soliton-like objects \cite{Tong:2003pz}--\cite{Eto:2005cp}  
as well as 1/8 BPS states \cite{Lee:2005sv,Eto:2005sw}, 
see \cite{Tong:2005un,Eto:2006pg} as a review.

     Solitons are solutions of non-linear partial differential equations (PDE). From mathematical point of view, an important question is whether these equations are integrable. One may take a different view and ask whether the solutions are unique in the sense of moduli space specification. 
     The integrability question of nonlinear PDE has been addressed to in a few different approaches. 
One, not rigorous, but practical means of analysis is the one which uses the Painlev\'e property of non-linear ordinary differential equation (ODE) and it's extension to PDE \cite{ARSc:1980}--\cite{Ward:1984gw}. 
This method was applied to many well known soliton equations. 
     From field theoretical points of view, more interesting applications would be those to the instanton equation 
and the monopole equation in Yang-Mills theory.
The application was made to self-dual Yang-Mills \cite{Jimbo:1982bg,Ward:1984gw} paralleled the ADHM construction of instantons. 
     The purpose of this paper is to study the integrability question (uniqueness question in other word) of the soliton equations in low dimensions which are derived from susy Yang-Mills theory in the Higgs phase, 
BPS vortex equation ($d=2+1$) and domain-wall equation ($d=1+1$).

     The Painlev\'e property is seldom used in particle physics, and we find it desirable to review it briefly, which we will do in section 2. 
We then introduce the model of gauge theories in the Higgs phase in section 3. 
In section 4 and 5 we examine the Painlev\'e property 
of domain walls and vortices, respectively. 
Section 6 is devoted to conclusion and discussion.

%%%%%%%%%%%%%%%%%%%%%%%%%%%%%%%%%%%%%%%%%%%%%%%%%%%%%%%%%

\section{Review of the Painlev\'e Test}

Given a nonlinear equation, 
the first question one should address to is its (non)integrability.
Integrability has been the notion for Hamiltonian systems of finite degrees of freedom. For the systems of $N$-degrees of freedom, if $N$ conserved quantities which are mutually involutive are found, the initial value problem can be solved by finite times of quadrature. 

If the notion of integrability is extended to systems of infinite degrees of freedom, it does not have precise definition, although some sufficient conditions, like inverse scattering transformation class, are known. There are a few variations of it, which depend on the system in question. However, it has not been proven that they are equivalent to each other. The Painlev\'e property is one of such definitions, which we will use in this paper. We call the equation which has Painlev\'e property P-type.

A non-linear ordinary differential equation is said to have 
the Painlev\'e property 
if it has the solutions whose movable singularities, 
depending on the initial condition, 
are only poles. 
Regarding the ordinary differential equation (ODE for short), if it has the Painlev\'e property, its solution can be expanded in the Laurent series near the movable singularity. The analysis of the Painlev\'e test for ODE uses this property.

Regarding the partial differential equation (PDE), there is a conjecture, known as Ablowitz-Ramani-Segur conjecture (ARS conjecture), connecting its integrability to Painlev\'e property for ODE \cite{ARSc:1980}--\cite{Ward:1984gw}. 
It says that every nonlinear ODE obtained by exact reduction of a nonlinear PDE  which is integrable by inverse scattering method is of P-type. 
There are cases in which the ODE obtained in this way would not be of P-type at first look, even if the PDE is integrable. So we have to find certain transformation of variables which makes the ODE to pass the Painlev\'e test. This is not a convenient way to check whether the PDE is integrable.

Weiss, Tabor and Carnevale proposed an analytical way to study integrability of  PDE \cite{WTC:1983}, called Weiss-Tabor-Carnevale analysis. 
They have introduced the notion of integrable PDE in the following sense. 
If the PDE is integrable, it has a singularity manifold. 
It is written by
\begin{eqnarray}
\phi(z_1,z_2,\cdots,z_n)=0\quad,\quad(\phi:\mbox{arbitrary function},\ \phi_{z_i}\neq 0).
\end {eqnarray}
We can expand integrable PDE near the singular manifold, in an analogous way that ODE can be expanded in a Laurent series near the movable singular point. 
We use this method to examine whether our models are integrable.

%%%%%%%%%%%%%%%%%%%%%%%%%%%%%%%%%%%%%%%%%%%%%%%%%%%%%%
%The Model

%\label{l5+1} "lagrangian in 5+1 dim" (\ref{l5+1})

%%%%%%%%%%%%%%%%%%%%%%%%%%%%%%%%%%%%%%%%%%%%%%%%%%%%%%%

\section{The Model and Soliton Equations}

\subsection{The Model and the Self-dual Yang-Mills-Higgs Equations}

We consider ${\cal N}=1$ supersymmetric Yang-Mills theory coupled with Higgs fields (Yang-Mills-Higgs system) in 5+1 dimensions. 
We limit our discussion to static solutions throughout the paper. 
We obtain Lagrangians in lower dimensions 
and their solutions 
by dimensional reduction of this system. 
Imposing the BPS condition on them, we obtain
the instanton-like equation in 4+1 dimensions \cite{Hanany:2004ea,Eto:2004rz}, 
the monopole-like equation in 3+1 \cite{Tong:2003pz}--\cite{Isozumi:2004vg} , 
the wall web equation in 3+1 \cite{Eto:2005cp}, 
the vortex equation in 2+1 \cite{ANO},\cite{Hanany:2003hp}--\cite{Eto:2006uw}, 
and the domain-wall equation 
in 1+1 \cite{Tong:2002hi,Isozumi:2004jc,Eto:2006uw}. 

In this paper 
we consider the $SU(N_{\rm C}) \times U(1)$ gauge theory coupled to 
$N_{\rm F}$ hypermultiplets as matter (Higgs) fields 
which are in the fundamental representation 
and constitute $SU(2)_R$ doublet:  
the theory consists of a gauge field $W_M$ ($M=0,1,2,3,4,5$) 
and Higgs fields $H^i$ $(i=1,2)$ in 
the form of $N_{\rm C}$ by $N_{\rm F}$ matrices. 
Supersymmetry forbids masses for hypermultiplets in $d=5+1$ and 
there exist the $SU(N_{\rm F})$ flavor symmetry. 
We will be concerned with the bosonic part, 
and we set the fermion fields to zero.  
The Lagrangian for the bosonic sector in $d=5+1$ is given by
\begin{eqnarray}
\L_{5+1} = \Tr\Bigl[-\frac{1}{2g^2}F^{MN}F_{MN}+\D^MH^i(\D_MH^i)^{\dagger}\Bigr]-V,\label{l5+1}
\end{eqnarray}
where the gauge field strength and covariant derivative are given by
\begin{eqnarray}
& F_{MN}=-i[\D_M,\D_N]=\partial_M W_N-\partial_N W_M+i[W_M,W_N] ,& \\
& \D_MH^i=\partial_M H^i+iW_MH^i, &
\end{eqnarray} 
and the potential $V(H^i)$ is given by
\begin{eqnarray}
 V = \frac{g^2}{4}\Tr\Bigl[(c^a {\bf 1}_{N_{\rm C}}-(\sigma^a)^j_i H^i(H^j)^{\dagger})^2\Bigr] 
\quad,\quad
(a=1,2,3).
\end{eqnarray}
Without loss of generality
we may choose the Fayet-Iliopoulos parameter as 
$c^a=(0,0,c)$ with $c>0$ by using $SU(2)_R$. 
Nonzero value of $c$ breaks $SU(2)_R$ explicitly. 

We consider static configuration. After dimensional reduction along $x^5$ and we ignore $W^5$. 
The second Higgs field $H^2$ vanishes for soliton configurations 
considered in this paper. 
We set $H \equiv H^1$. The potential reduces to  
\begin{eqnarray}
 V = \frac{g^2}{4}\Tr\Bigl[(c{\bf 1}_{N_{\rm C}} - H H^{\dagger})^2\Bigr] .
\end{eqnarray} 
The energy density for static configuration ($\partial_{x^0} = 0$)
is written as follows \cite{Hanany:2004ea,Eto:2004rz}
\begin{eqnarray}
{\cal E} &=& \Tr\left[
\frac{1}{2g^2}F_{mn}F_{mn} 
+ {\cal D}_mH({\cal D}_mH)^\dagger
+ \frac{g^2}{4} (c{\bf 1}_{N_{\rm C}} - H H^{\dagger})^2
\right]\nonumber\\
&=& 
 \Tr\bigg[\frac{1}{g^2}\left\{
 \left(F_{13} - F_{24}\right)^2
 + \left(F_{14} + F_{23}\right)^2
 + \left(F_{12} + F_{34} + \frac{g^2}{2} (c{\bf 1}_{N_{\rm C}} - H H^{\dagger})
 \right)^2 
\right\} \nonumber\\
&& 
+({\cal D}_1 H + i  {\cal D}_2 H)^\dagger 
({\cal D}_1 H + i  {\cal D}_2 H) 
+({\cal D}_3 H + i  {\cal D}_4 H)^\dagger 
({\cal D}_3 H + i  {\cal D}_4 H)
\nonumber\\
&& + \frac{1}{2g^2}F_{mn}\tilde F_{mn} 
- c(F_{12} + F_{34})
+ \partial_m J_m
\bigg]
 \label{BPS-bound}
\end{eqnarray}
where $m,n=1,2,3,4$. 
The energy density is minimized if the following set of BPS equations 
is satisfied, 
\begin{eqnarray}
&&
F_{13}- F_{24} =0, \ \ 
F_{14} + F_{23} =0, \ \ 
\label{eq:4eq:ivv1}\\
&& F_{12} +  F_{34} 
  = - \frac{g^2}{2} (c{\bf 1}_{N_{\rm C}} - H H^{\dagger}), \ \ 
\label{eq:4eq:ivv2}\\
&&
{\cal D}_1 H + i  {\cal D}_2 H =0, \ \ 
{\cal D}_3 H + i  {\cal D}_4 H =0. 
\label{eq:4eq:ivv3}
\end{eqnarray}
If we set the Fayet-Iliopoulos parameter $c$ to zero and set $H=0$, 
we obtain the well known 
self-dual Yang-Mills (SDYM) equations for instantons. 
We call Eqs.~(\ref{eq:4eq:ivv1})--(\ref{eq:4eq:ivv3}) the SDYM-Higgs equations.
Several mathematicians have also studied these equations \cite{math}.\footnote{
The SDYM-Higgs equations (\ref{eq:4eq:ivv1})--(\ref{eq:4eq:ivv3}) 
in $d=4+0$ can be derived from 
the Donaldson-Uhlenbeck-Yau equations in $d=6+0$ \cite{DUY} 
by the $SU(2)$ equivariant dimensional reduction 
on $S^2$, at least in the case of $U(1)$ gauge group \cite{Popov:2005ik}. 
We thank A.~D.~Popov to point this out. 
} 
We will be concerned with the soliton equations in lower dimensions, $d<4+1$, which are obtained by dimensional reduction from 
Eqs.~(\ref{eq:4eq:ivv1})--(\ref{eq:4eq:ivv3}). 
Now, we are going to show vortices and domain walls in $d=2+1,1+1$, respectively.

%%%%%%%%%%%%%%%%%%%%%%%%%%%

\subsection{Vortex Equation}

We first perform the trivial dimensional reduction of (\ref{l5+1}) 
to three dimensions. 
We choose the coordinates which are reduced to be $x^3,x^4,x^5$. 
After dimensional reduction, gauge fields $W_3,W_4,W_5$ are adjoint scalars (in $d=3$) $\Sigma_{\alpha},(\alpha=1,2,3)$. 
In the following we set $\Sigma_{\alpha} = H^2 = 0$
because they do not contribute to vortex solutions (and set $H=H^1$).
The Lagrangian (\ref{l5+1}) is now written as
\begin{eqnarray}
\L_{2+1}=\Tr\left[-\frac{1}{2g^2}F_{mn}F^{mn}
 + \D_m H(\D^m H)^{\dagger}\right] 
- \frac{g^2}{4}\Tr[(c {\bf 1}_{N_{\rm C}} - H H^{\dagger})^2] ,
\end{eqnarray}
with $m=0,1,2$

We next consider the BPS states of our theory. 
We ignore the $x^0$ dependence. 
The energy density of static configurations is given by
\begin{eqnarray}
 \E &=& \Tr\Biggl[ \frac{1}{g^2}(F_{12})^2 
  + \{ \D_1H(\D_1H)^{\dagger} + \D_2H(\D_2H)^{\dagger} \}  
 + \frac{g^2}{4} (c {\bf 1}_{N_{\rm C}} - H H^{\dagger})^2 \Biggr]  \nonumber \\
 &=& {\rm Tr}\left[\frac1{g^2}
\left(F_{12} + {g^2 \over 2} (c {\bf 1}_{N_{\rm C}} - H H^\dagger)\right)^2
 +\left({\cal D}_1 H + i {\cal D}_2 H\right)
 \left({\cal D}_1 H + i {\cal D}_2 H\right)^\dagger\right]\nonumber \\
  &&+{\rm Tr}\left[-c\,F_{12} + 2i\partial _{[1}H{\cal D}_{2]}H^\dagger \right].
\end{eqnarray}
This energy is minimized if the following BPS equations 
for non-Abelian vortices are satisfied 
\cite{Hanany:2003hp,Auzzi:2003fs,Eto:2005yh,Eto:2006pg}: 
\begin{eqnarray}
 & 0=\D_1H+i\D_2H , &
   \label{vortex-eq1} \\
 & 0=F_{12}+\frac{g^2}{2}(c{\bf 1}_{N_{\rm C}}-HH^{\dagger}) . &
   \label{vortex-eq2}
\end{eqnarray}
These equations are called the vortex equations. 
These can also be obtained by ignoring $x^3$ and $x^4$ dependence 
and setting $W_3 = W_4 =0$  
in Eqs.~(\ref{eq:4eq:ivv1})--(\ref{eq:4eq:ivv3}).  
In the case of $U(1)$ gauge theory with a single Higgs field 
and hence $U(1)$ flavor symmetry alone, 
this BPS state is known as Abrikosov-Nielsen-Olesen (ANO) 
vortex \cite{ANO}.

Following the same line as the Yang's equation \cite{Yang:1977zf} 
for self-dual Yang-Mills fields, 
we rewrite Eqs.~(\ref{vortex-eq1}) and (\ref{vortex-eq2}) 
in the second order PDE \cite{Eto:2005yh,Eto:2006mz,Eto:2006uw,Eto:2006pg}. 
First we define complex notations 
\begin{eqnarray}
 z \equiv x^1 + i x^2 , \quad 
 \Wb_z\equiv\frac{W_1+iW_2}{2}. 
\end{eqnarray}
Then note the first vortex equation (\ref{vortex-eq1}) can be integrated as
\begin{eqnarray}
\Wb_z = -iS^{-1}\parb_zS,
\quad
 H = S^{-1}H_0(z),
\end{eqnarray}
where we have introduced $S = S(z,z^*) \in GL(N_{\rm C},{\bf C})$ 
and $H_0(z)$ is an $N_{\rm C} \times N_{\rm F}$ matrix whose components 
are holomorphic with respect to $z$. 
Constants in $H_0$ are integration constants as moduli.
Defining a gauge invariant
\begin{eqnarray}
\Omega(z,z^*) \equiv SS^\dagger,
\end{eqnarray}
the second vortex equation (\ref{vortex-eq1}) can be rewritten as 
\begin{eqnarray}
 \partial_z(\Omega^{-1}\parb_z\Omega) 
 = \frac{g^2}{4}(c{\bf 1}_{N_{\rm C}}-\Omega^{-1}H_0H_0^\dagger).
  \label{master-eq-vortex}
\end{eqnarray}
We call this equation {\it the master equation} for vortices.
We will examine if this equation has the Painlev\'e property  
in section \ref{sec:PTV}.

%%%%%%%%%%%%%%%%%%%%%%%%%%%%%%%%%%%%%%%%%%%%%%%%%%%%

%\label{medw} "master equation domain walls" (\ref{medw})

%%%%%%%%%%%%%%%%%%%%%%%%%%%%%%%%%%%%%%%%%%%%%%%%%%%%5

\subsection{Domain-Wall Equation}

To obtain the domain-wall equation we have to give a mass 
to the Higgs fields $H$. This can be made by using the Scherk-Schwarz dimensional reduction \cite{Scherk:1979zr} with respect the coordinate $x^2$, after the simple reduction to three dimensions of the previous sector.

We are considering Scherk-Schwarz dimensional reduction to a certain theory in $2+1$ dimension; the coordinates are $x^0,x^1,x^2$. We would like to reduce one of those; we set it $x^2$. The dependence of the Higgs field 
on it will be compactified to $S^1$ by using twisted boundary condition
\begin{eqnarray}
H(x^{\mu},x^2+2\pi R)=H(x^{\mu},x^2)e^{i2\pi RM}
\end{eqnarray}
with $\mu=0,1$ and 
\begin{eqnarray*}
 M = {\rm diag.}(m_1,m_2,\cdots,m_{N_{\rm F}}) , \quad
 0 \leq m_A < 1/R \quad , \quad (A=1,2,\cdots,N_{\rm F}). &
\end{eqnarray*}
We take the lowest massive mode ignoring the infinite tower of 
higher Kaluza-Klein modes.
We write
\begin{eqnarray}
 H (x^{\mu},x^2)=\frac{1}{\sqrt{2\pi R}}\hat{H} (x^{\mu})e^{iMx^2}.
\end{eqnarray}
The other fields do not depend on the coordinate $x^2$:
\begin{eqnarray}
\nonumber
& W_{\mu}(x^{\mu},x^2)=W_{\mu}(x^{\mu}), \quad
 \Sigma(x^{\mu},x^2)=\Sigma(x^{\mu}), \quad
 W_2(x^{\mu},x^2)=-\hat{\Sigma}(x^{\mu}). 
\end{eqnarray}
After the Lagrangian (\ref{l5+1}) is reduced by trivial dimensional reduction 
and Scherk-Schwarz dimensional reduction, 
we can write the Lagrangian in $1+1$ dimensions as
\begin{eqnarray}
\L_{1+1}=\Tr\Bigl[-\frac{1}{2g^2}\bigl\{F_{\mu\nu}F^{\mu\nu}-2\D_{\mu}\hat{\Sigma}\D^{\mu}\hat{\Sigma}\bigr\}+\D_{\mu}H (\D^{\mu}H)^{\dagger}\Bigr]-V,
\end{eqnarray}
where the potential is given by 
\begin{eqnarray}
\nonumber
& V = \frac{g^2}{4} 
  \Tr\Bigl[-\bigl(c {\bf 1}_{\Nc}-H H^{\dagger}\bigr)^2 \Bigr] 
 + \Tr\Bigl[(\hat{\Sigma}H -H M)(\hat{\Sigma}H -H M)^{\dagger}\Bigr]. &
\end{eqnarray}
Here, the covariant derivative of $\hat{\Sigma}$ is written by
\begin{eqnarray}
\D_m \hat \Sigma = \partial_m \hat \Sigma+i[W_m,\hat \Sigma].
\end{eqnarray}
The vacua of the theory are determined as the minimum of the potential $V$: 
\begin{eqnarray}
\nonumber
 H H^\dagger = c {\bf 1}_{\Nc} , \quad
 \hat{\Sigma}H - H M=0.  
\end{eqnarray} 
These can be solved as
\begin{eqnarray}
 && 
 H=\sqrt{c} \left(
  \begin{array}{ccccccccccc}
  0 & \cdots & 0 & 1 & 0 \cdots & 0 & 0 \cdots & & & & 0 \cdots\\
  \vdots &  & \vdots & 0 & \vdots & 1 & \vdots & & & & \vdots \\
  \vdots & & \vdots & \vdots & \vdots & \vdots & \vdots & 1 & & & \vdots\\
  \vdots & & \vdots & \vdots & \vdots & \vdots & \vdots & & \ddots&&\vdots \\
  0 & \cdots & 0 & 0 & 0 \cdots & 0 & 0 \cdots & & & 1 & 0 \cdots \\
        \end{array}\right) , \\
 && \hat \Sigma=\mbox{diag}(m_{A_1},m_{A_2},\cdots,m_{A_{N_{\rm C}}}). 
\end{eqnarray}
The first form is constructed by $\Nc$ unit vectors and $\Nf-\Nc$ zero vectors. Let us label the position of the unit vector as 
$[A_{N_1},A_{N_2},\cdots,A_{N_{\Nc}}]$.  And the disposition of the unit vectors can be chosen arbitrarily, except to be satisfied $A_r<A_{r+1}$ ($r=1,2,\cdots,N_{\rm C}$). 
Therefore the number of the vacua is $~_{N_{\rm F}} C_{N_{\rm C}}$.

We next consider the BPS states of our theory. 
Their energy density is given by 
\begin{eqnarray}
{\cal E} 
 &=& {\rm Tr} 
  \left[ {1 \over g^2} ({\cal  D}_y \hat \Sigma)^2 
   + {\cal D}_y H {\cal D}_y H^\dagger
   + \frac{g^2}{4} \bigl(c {\bf 1}_{\Nc}-H H^{\dagger}\bigr)^2 
  + (\hat{\Sigma}H -H M)(\hat{\Sigma}H -H M)^{\dagger} \right] \nonumber \\
 &=& {1 \over g^2}{\rm Tr}\left[{\cal  D}_y \hat \Sigma -
 {g^2\over 2}\left(c{\bf 1}_{N_{\rm C}} - H H^\dagger 
 \right)\right]^2
 \nonumber \\
 &&{}
 + {\rm Tr}\left[
 ({\cal D}_y H + \hat \Sigma H - H M) 
 ({\cal D}_y H + \hat \Sigma H - H M)^\dagger\right] 
 \nonumber \\ && {}
 + c \,\partial_y{\rm Tr}\hat \Sigma 
 - \partial_y \left\{{\rm Tr}
 \left[ 
 \left(\hat \Sigma H - H M\right)H^\dagger
 \right]\right\}, 
 \label{eq:wll:bogomolnyi-wall}
\end{eqnarray} 
with $y \equiv x^1$. 
The energy density is minimized 
by imposing the BPS equations of domain walls (domain-wall equation) 
\begin{eqnarray}
 & \D_yH = - \hat \Sigma H + H M , & \label{wall-eq1} \\
 & \D_y \hat \Sigma = {g^2 \over 2}(c{\bf 1}_{\Nc}-H H^{\dagger}) .& 
  \label{wall-eq2} 
\end{eqnarray}
These equations can be directly derived from vortex equations 
(\ref{vortex-eq1}) and (\ref{vortex-eq2}) by 
the Scherk-Schwarz dimensional reduction 
$H(x^0,x^1,x^2) \to H(x^0,x^1)e^{i M x^2}$. 

We rewrite the set of domain-wall equations 
to a second order ODE in the form  
like the Yang's equation \cite{Isozumi:2004jc,Eto:2006pg}.  
The first of the domain-wall equation (\ref{wall-eq1}) can be integrated as  
\begin{eqnarray}
 H = S^{-1}H_0e^{My} , \\
 \hat{\Sigma}+iW_y=S^{-1}\partial_yS ,
\end{eqnarray}
with an $N_{\rm C} \times N_{\rm F}$ matrix $H_0$ 
of integration constants, 
and $S = S(y) \in GL(N_{\rm C},{\bf C})$.  
Defining a gauge invariant
\begin{eqnarray}
 \Omega (y) \equiv SS^{\dagger}, 
\end{eqnarray}
the second of the domain-wall equation (\ref{wall-eq2}) 
can be rewritten to the equation which we want: 
\begin{eqnarray}
\partial_y(\Omega^{-1}\partial_y\Omega)
 = g^2(c{\bf 1}_{N_{\rm C}}-\Omega^{-1}H_0e^{2My}(H_0)^{\dagger}).
  \label{medw}
\end{eqnarray}
We call this equation {\it the master equation} for domain walls.
We will examine if this equation has the Painlev\'e property  
in the next section.

%%%%%%%%%%%%%%%%%%%%%%%%%%%%%%%%%%%%%%%%%%%%%%%%%%%
%Painlev\'e Test for $U(1),N_f=1$ Domain Walls

%\label{cfmedw1} "coupled function master equation domain wall 1" (\ref{cfmedw1})
%\label{cfmedw2} "coupled function master equation domain wall 2" (\ref{cfmedw2})
%\label{eeo} "extended equation ODE" (\ref{eeo})

%%%%%%%%%%%%%%%%%%%%%%%%%%%%%%%%%%%%%%%%%%%%%%%%%%%%

\section{Painlev\'e Test for Domain Walls}

In this section, we will check whether the master equation (\ref{medw}) 
of domain wall has Painlev\'e property. 
Since this master equation is the ODE, 
we can do it easier than PDE like the case of vortices. 

From now on we consider the simple case of 
$N_{\rm C}=1,N_{\rm F}=2$. 
This is the simplest case admitting a single domain wall \cite{Abraham:1992vb}.
We set mass $M$ to be 
\begin{eqnarray}
 M =  \left(
     \begin{array}{cc}
       0  & {} \\
       {} & -\Delta m \\
     \end{array}\right)
\end{eqnarray}
and put the moduli matrix 
\begin{eqnarray}
 H_0 = \sqrt{c}(1,e^{\Delta m y_0}) 
\end{eqnarray}
where $y_0$ is a constant.
Defining $\psi$ by 
\begin{eqnarray}
& \Omega \equiv e^{2\psi(y)} 
\end{eqnarray}
the master equation (\ref{medw}) can be rewritten 
as \cite{Tong:2002hi,Isozumi:2003rp}
\begin{eqnarray}
 \frac{d^2\psi}{dy^2}=\frac{g^2c}{2}\{1-e^{-2\psi}(e^{-2\Delta m(y-y_0)}+1)\}.
\end{eqnarray} 

This equation has the factor $e^{-2\psi}$ which has to be expanded 
in a power series of $\psi$. 
It makes the Painlev\'e test complicated. 
Therefore we transform this equation into two coupled equations: 
\begin{eqnarray}
& \displaystyle
   w \frac{dA}{dw}=AB , &\label{cfmedw1} \\
& \displaystyle
   w \frac{dB}{dw}=\frac{1}{2k^2}\{A(w+1)-1\}\label{cfmedw2} ,&
\end{eqnarray}
where we have defined
\begin{eqnarray}
& w \equiv \exp (-2\Delta m(y-y_0)), & \\
& \displaystyle
    A(w) \equiv \exp (-2\psi(y))\quad , \quad
    B(w) \equiv \frac{\psi '(y)}{\Delta m}, & \\
& \displaystyle
    k \equiv \sqrt{\frac{2(\Delta m)^2}{g^2c}}. &
\end{eqnarray}
The ODE (\ref{cfmedw1}) and (\ref{cfmedw1}) have now only one 
dimensionless parameter $k$. 
The variable $y$ takes a value in ${\bf R}$ but 
we now consider $w$ to take a value in ${\bf C}$ by
analytic continuation. 

We assume that the solutions $A(w)$ and $B(w)$ have 
a movable singularity $w=w_*$. 
We then expand $A(w)$ and $B(w)$ in the Laurent series 
near $w_*$:
\begin{eqnarray}
A(w)=(w-w_*)^{-\alpha}\sum_{l=0}^{\infty}A_l(w-w_*)^l \quad,\quad
B(w)=(w-w_*)^{-\beta}\sum_{l=0}^{\infty}B_l(w-w_*)^l.\label{eeo}
\end{eqnarray}
Here $\alpha$ and $\beta$ are positive constants and we assume that
\begin{eqnarray}
A_0 \neq 0\quad ,\quad B_0 \neq 0 \quad \mbox{and} \quad A_l=B_l=0 
 \quad(l<0).
\end{eqnarray} 
We substitute the expansion (\ref{eeo}) into 
Eqs.~(\ref{cfmedw1}) and (\ref{cfmedw2}). 
%%%%%%%%%%%%%%
The leading order analysis yields
\begin{eqnarray}
 \nonumber
& \alpha=2 \quad,\quad \beta=1,& \\
& \displaystyle
   A_0=4k^2\frac{w_*^2}{w_*+1} \quad,\quad B_0=-2w_*. &\label{rfdw} 
\end{eqnarray}
We also obtain the recursive relations 
\begin{eqnarray}
 \nonumber
& \displaystyle (l-3)A_{l-1}+w_*(l-2)A_l=\sum_{j=0}^lA_{l-j}B_j , &
\\
& \displaystyle (l-2)B_{l-1}+w_*(l-1)B_l=\frac{1}{2k^2}(w_*+1)A_l
 -\frac{1}{2k^2}\delta_{n,2} .& \label{recursive0} 
\end{eqnarray} 
We now examine from these recursive relations  
whether certain degrees $l$ can become resonances, or not. 
The relations (\ref{recursive0}) can be 
written as the following form: 
\begin{eqnarray}
(l+1)(l-2)w_*B_l=F_l(A_{l-1},\ldots,A_0,B_{l-1},\ldots,B_0) , 
    \label{recursive}
\end{eqnarray}
with some function $F_l$.  
We find that the degrees which are possible to be resonances are $l=-1,2$ 
because $B_l$ (or $w_*$) can become arbitrary in these cases. 
The one of possibilities of resonance, $l=-1$, 
comes from the arbitrariness of $w_*$. 
If there would exist another resonance in 
Eqs.~(\ref{cfmedw1}) and (\ref{cfmedw2}), 
it could be $l=2$.   
In this case, $F_2$ must vanish from (\ref{recursive}). 
However $F_2$ is obtained from (\ref{recursive0}) as
\begin{eqnarray}
F_2=\frac{w_*}{(w_*+1)^2}-\frac{1}{2k^2}, 
\end{eqnarray}
and this is impossible to vanish 
due to the arbitrariness of $w_*$.  
Therefore $l=2$ cannot become a resonance.

From these analyses, we have seen that there are no resonances. 
Therefore the master equation  (\ref{medw}) for domain walls 
has no Painlev\'e property in the case of $N_{\rm F}=2$, $N_{\rm C}=1$. 
Since $U(N_{\rm C})$ gauge theory with $N_{\rm F} (>N_{\rm C})$ flavors 
contains the $N_{\rm F}=2$, $N_{\rm C} =1$ case discussed above
we conclude that 
the master equation (\ref{medw}) for domain walls 
has no Painlev\'e property in general cases. 
This implies that domain-wall equations are not integrable.

%%%%%%%%%%%%%%%%%%%%%%%%%%%%%%%%%%%%%%%%%%%%%%%%%%%

%%{Painlev\'e Test for $U(1),N_f=1$ Vortices}

%\label{ymev} "yang's master equation vortex" (\ref{ymev})
%\label{cemev} "coupled equations master equation vortex" (\ref{cemev})
%\label{eep} 'expanded equations PDE' (\ref{eeop})
%\label{rfv} "recursive functions vortex" (\ref{rfv})

%%%%%%%%%%%%%%%%%%%%%%%%%%%%%%%%%%%%%%%%%%%%%%%%%%%%5

\section{Painlev\'e Test for Vortices} \label{sec:PTV}

In this section, we will check integrability of 
the master equation (\ref{master-eq-vortex}) for vortices. 
Since the master equation (\ref{medw}) for domain wall is not integrable 
and it is obtained from the master equation (\ref{master-eq-vortex}) 
of vortices by dimensional reduction, 
it is likely to be non-integrable 
from the ARS conjecture \cite{ARSc:1980}. 
We will see that this is the case by directly performing 
the Painlev\'e test of the master equation (\ref{master-eq-vortex}) 
for vortices.

We consider the case of
$N_{\rm C}=1,N_{\rm F}=1$,  
the simplest case of the ANO vortices \cite{ANO}.
We again define a new variable as
\begin{eqnarray}
& \displaystyle
\Omega=e^{\psi(z,\bar{z})} . & 
\end{eqnarray}
We consider the single vortex case:
\begin{eqnarray}
 & H_0=\sqrt{c}(z-z_1). &
\end{eqnarray}
If there were $n$ vortices, $H_0$ should become 
$H_0=\sqrt{c}\Pi^n_{k=1}(z-z_k)$ \cite{Eto:2005yh}, 
but we can repeat the following analysis in the same way. 
Then, the master equation (\ref{master-eq-vortex}) is rewritten as
\begin{eqnarray}
\partial_z\partial_{\bar{z}}\psi 
= \frac{g^2}{4}\{c-e^{-\psi}(z-z_1)(\bar{z}-\bar{z}_1)\}.\label{ymev}
\end{eqnarray}

It is impossible to check the integrability by the Painlev\'e test, 
when there exist exponential terms which depend on $z$ in PDE. 
In order to do it, 
we need to change the form of equation (\ref{ymev}) again. 
The variables are changed as
\begin{eqnarray}
\nonumber
& A(z,\bar{z})\equiv \exp[-\psi(z,\bar{z})] ,& \\
& B(z,\bar{z})\equiv \partial_{\bar{z}}\psi(z,\bar{z}) . &\label{eep}
\end{eqnarray}
Then we obtain the two coupled equations, given by 
\begin{eqnarray}
& \displaystyle
\partial_{\bar{z}}A= -AB , & \label{mevor1} \\
& \displaystyle
\partial_zB=\frac{g^2}{4}\bigl\{c-A(z-z_1)(\bar{z}-\bar{z}_1)\bigr\} .&
  \label{cemev}
\end{eqnarray}

We assume that the solutions $A(z,\bar{z})$ and $B(z,\bar{z})$ can be expanded near a singularity manifold, given by $\phi(z,\bar{z})=0$, as 
\begin{eqnarray}
 A(z,\bar{z})=\sum_{l=0}^{\infty}A_l(z,\bar{z})\phi^{l-m}(z,\bar{z})
\quad , \quad
 B(z,\bar{z})=\sum_{l=0}^{\infty}B_l(z,\bar{z})\phi^{l-n}(z,\bar{z}), 
  \label{vor-expansion}
\end{eqnarray}
where, $m$ and $n$ are positive constants, and we assume that
\begin{eqnarray}
A_0\neq0 \ ,\ B_0\neq0 \quad \mbox{and}\quad A_l=B_l=0 \quad (l<0).
\end{eqnarray}
There is the element $(z-z_1)(\bar{z}-\bar{z}_1)$ in Eq.~(\ref{cemev}). 
This makes a difference of coefficient functions between a case 
(the case 1) that the function $\phi$, 
defining the singularity manifold, 
does not have zeros at $z=z_1$ 
($\phi(z=z_1) \neq 0$)
and 
another case (the case 2) 
that it has zeros at least of first order at $z=z_1$ 
($\phi(z=z_1) = 0$). 
In order to identify each other, 
we will write those singularity manifolds as 
$\phi_1=0$ and $\phi_2=0$, respectively. 
Also we check these Painlev\'e property separately.

\medskip

\noindent
\underline{The case 1}. 
First we will consider the case in which the singularity manifold 
is given by $\phi_1=0$. 
If we substitute the expansion (\ref{vor-expansion}) into 
Eqs.~(\ref{mevor1}) and (\ref{cemev}), 
the leading order analysis yields
\begin{eqnarray}
\nonumber
& m=2 \quad , \quad n=1, & \\
& \displaystyle
A_0=\frac{8}{g^2}\frac{(\partial_{\bar{z}}\phi_1)(\partial_z\phi_1)}{(z-z_1)(\bar{z}-\bar{z}_1)} \quad , \quad B_0=2(\partial_{\bar{z}}\phi_1).  &
\end{eqnarray}

\noindent
We also obtain the recursive functions 
\begin{eqnarray}
\nonumber
& \displaystyle
\partial_{\bar{z}}A_{l-1}+(l-2)A_l(\partial_{\bar{z}}\phi_1)=-\sum_{j=0}^{l}A_{l-j}B_j\quad(l\geq 1) , & \\
& \displaystyle
\partial_{z}B_{l-1}+(l-1)B_l(\partial_z\phi_1)=-\frac{g^2}{4}(z-z_1)(\bar{z}-\bar{z}_1)A_l+\frac{g^2}{4}c\delta_{l,2} \quad (l\geq 1) .&\label{rfv}
\end{eqnarray}
We can find certain degrees $l$ which are possible to become resonances 
from above recursive functions (\ref{rfv}). 
Those functions can be written as  
\begin{eqnarray}
\frac{4}{g^2}(l-2)(l+1)
  \frac{(\partial_z\phi_1)(\partial_{\bar{z}}\phi_1)}
       {(z-z_1) ({\bar{z}-\bar z_1})} B_l 
  = F_l(A_{l-1},\ldots,A_0,B_{l-1},\ldots,B_0) 
  \label{F}
\end{eqnarray}
with some function $F_l$.  
Now we find from the left hand side that 
the degrees which are possible to be resonances are $l=-1,2$,
because $B_l$ (or $\phi_1$) 
can be arbitrary function in these cases. 
The one of possibilities of resonance, $l=-1$, 
comes from the arbitrariness of $\phi_1$. 
If there exists another resonance in Eqs.~(\ref{mevor1}) and (\ref{cemev}), 
it should be $l=2$. 
Therefore we now check whether it is the resonance, 
by using the recursive relations (\ref{rfv}). 
If we think of the case $l=1$ for Eq.~(\ref{rfv}), 
we obtain precise expression of $A_1$ and $B_1$:  
\begin{eqnarray}
\nonumber
& \displaystyle
 A_1=\frac{A_0B_1+\partial_zA_0}{(\partial_{\bar{z}}\phi_1)-B_0}, & \\
 \nonumber
 & \displaystyle
 B_1 = -\Bigl\{\frac{g^2}{4}(z-z_1)(\bar{z}-\bar{z}_1)
              \frac{A_0}{\partial_{\bar{z}}\phi_1} 
          -(\partial_z\phi_1)\Bigr\}^{-1}
      \Bigl\{\partial_zB_0+\frac{g^2}{4}(z-z_1)
      (\bar{z}-\bar{z}_1)\frac{\partial_{\bar{z}}A_0}
      {(\partial_{\bar{z}}\phi_1)-B_0}\Bigr\} & \\
 & \displaystyle
  {} =-\Bigl\{2\frac{(\partial_{\bar{z}}^2\phi_1)(\partial_z \phi_1)}
  {\partial_{\bar{z}}\phi_1}+\frac{(\partial_{\bar{z}}\phi_1)
    (\partial_z \phi_1)}{\bar{z}-\bar{z}_1}\Bigr\} . &
\end{eqnarray} 
For $l=2$, the left hand side of (\ref{F}) vanishes while 
we can show from these equations and the recursive functions (\ref{rfv})
that the right hand side of (\ref{F}) does not vanish, $F_2 \neq 0$, 
like the case of domain walls. 
It means that $l=2$ is not a resonance. 
Hence, Eqs.~(\ref{mevor1}) and (\ref{cemev}) have no resonances 
in the case 1.

%\vspace{2cm}
\medskip
\noindent
\underline{The case 2}.
Next we will consider the case that the singularity manifold is 
given by $\phi_2=0$. 
We write this singularity manifold as
\begin{eqnarray}
 M \equiv \frac{\phi_2}{(z-z_1)(\bar{z}-\bar{z}_1)}
\end{eqnarray}
with a function $M(z,\bar z)$,
in order to clarify 
that the coefficient functions do not have any singularities.
Here, $M$ could depend on $(z-z_1)(\bar{z}-\bar{z}_1)$. 
If we substitute (\ref{vor-expansion}) 
into Eqs.~(\ref{mevor1}) and (\ref{cemev}), 
the leading order analysis yields 
\begin{eqnarray}
\nonumber
& \displaystyle
m=3 \quad , \quad n=1, & \\
& \displaystyle
 A_0=\frac{12}{g^2}M(\partial_z\phi_2) (\partial_{\bar{z}}\phi_2) 
  \quad , \quad 
 B_0=3(\partial_{\bar{z}}\phi_2). &
\end{eqnarray}
We also obtain the recursive functions 
\begin{eqnarray}
\nonumber
& \displaystyle
\partial_{\bar{z}}A_{l-1}+(l-3)A_l(\partial_{\bar{z}}\phi_2) 
 = -\sum_{j=0}^lA_{l-j}B_j \quad (l\geq 1),& \\
& \displaystyle
 \partial_zB_{l-1}+(l-1)B_l(\partial_z\phi_2) 
 = \frac{g^2}{4}c\delta_{l,2}-\frac{g^2}{4}\frac{1}{M}A_l \quad (l\geq 1). 
  \label{rfv2}&
\end{eqnarray} 
We are concerned with whether there are some possibilities of resonance, or not. We check it by the same method as above. 
Then from (\ref{rfv2}) we get  
\begin{eqnarray}
\frac{4}{g^2}M(l^2-l-3)(\partial_{z}\phi_2)(\partial_{\bar{z}}\phi_2)B_l=F_l(A_{l-1},\cdots,A_0,B_{l-1},\cdots,B_0).
\end{eqnarray}
Therefore $B_l$ can be an arbitrary function only when 
$l=\frac{1\pm\sqrt{13}}{2}$. 
However the resonance must be an integer or at most rational number, 
so it is impossible that this number $l$ becomes a resonance. 
Therefore Eqs.~(\ref{mevor1}) and (\ref{cemev}) have no resonances.

From the above consideration, the cases 1 and 2, 
we conclude that the master equation (\ref{master-eq-vortex}) 
for $U(1)$, $\Nf=1$ vortex 
does not have the Painlev\'e property. 
Since equations for the theory with gauge group $U(N_{\rm C})$ 
and general number of flavor $N_{\rm F} (>N_{\rm C})$ contain 
this case at least, 
we conclude that the master equation (\ref{master-eq-vortex}) 
for vortices does not have the Painlev\'e property in general.
This fact implies that vortex equations are not integrable.

%%%  (end of vortex painleve test)  %%%

%%%%%%%%%%%%%%%%%%%%%%%%%%%%%%%%%%%%%%%%%%%
\section{Conclusions and Discussion}

The first (Higgs part) equations of vortex equation (\ref{vortex-eq1}) 
and domain-wall equation (\ref{wall-eq1}) can be integrated 
(to give integration constants $H_0$)
while the second 
(gauge part) of them, (\ref{vortex-eq2}) and (\ref{wall-eq2}), 
are rewritten into 
the master equations (\ref{master-eq-vortex}) and (\ref{medw}) 
for vortices and domain walls, respectively. 
We have shown that the master equations for 
domain walls (\ref{medw}) and vortices (\ref{master-eq-vortex})
do not pass the Painlev\'e test and therefore that they do not have 
the Painlev\'e property.
Of course there remains a possibility that 
better choice of variables might make the form of equations 
suitable to pass the Painlev\'e test.
However we believe that it is not the case from many trials. 
Our results imply that 
the domain-wall equation and the vortex equation are not integrable, 
because most equations which do not pass Painlev\'e test are not integrable. 
This is in contrast to the cases of instantons and monopoles, whose equations are integrable and have Painlev\'e property.

Our results are related with uniqueness problem of moduli. 
Solution of the master equation of vortices 
in the case of $N_{\rm C}=N_{\rm F}=1$
(equivalent to the Taubes equation) 
was shown to exist uniquely \cite{Taubes:1979tm}.\footnote{The uniqueness of solution of the master equation of arbitrary gauge group with arbitrary matter contents was also shown in the case that the base space is compact K\"ahler instead of ${\bf R}^2$ \cite{math}.} 
This implies that the master equation (\ref{master-eq-vortex}) 
does not contain moduli and therefore 
is not integrable. 
Thus all moduli are contained in $H_0(z)$. 
The uniqueness and existence of solution of 
the master equation of domain walls for $N_{\rm C}=1$ 
has also been proved recently \cite{Sakai:2005kz}.

Ward conjectured that 
all integrable equations are obtained 
from self-dual Yang-Mills (SDYM) equation \cite{Ward:1985gz} 
by some dimensional reductions. 
If this conjecture is true, the vortex equation and 
the domain-wall equation
should not be integrable, 
because they are obtained as dimensional reduction 
of the SDYM coupled with the Higgs fields 
(\ref{eq:4eq:ivv1})--(\ref{eq:4eq:ivv3}).
In this sense our results support the Ward's conjecture. 
Non-commutative version \cite{Hamanaka:2002mz} 
of our analysis is also interesting to explore.
In particular, the large non-commutativity limit may reduce equations integrable especially for vortices.

%==========================================
% Acknowledgements
%==========================================
\section*{Acknowledgements}
We would like to thank 
Ed Corrigan, 
Shun'ya Mizoguchi, 
Keisuke Ohashi, Alexander D.~Popov, 
Koichi Toda 
for valuable comments. 
T.~I. acknowledges supports from JSPS grants, Kiban C and Kiban B. 
The work of M.~N. was 
supported by JSPS
%Japan Society for the Promotion of Science 
under the Post-doctoral Research 
Program. 
T.~I. and S.~M. would like to thank C.~S.~Chu and Anne Taormina 
for the hospitality at Mathematical Science Department of University of Durham, where a part of the present work was done. 

%%%%%%%%%%%%%%%%%%%%%%%%%%%%%%%%%%%%%%%%%%%%%%%%%%%%%%%%%%%%%%%%%%%%

\end{document}